# `A Generalized Mechanical Model for the Cycle Rank Dependence of Stretch at Break in Phantom Chain Star Polymer Networks

Yuichi Masubuchi*, Takato Ishida, and Takashi Uneyama

*Department of Materials Physics, Nagoya University, Nagoya 464-8603, Japan*

To be submitted to Polymer, ver 8, Jul, 2026

* To whom correspondence should be addressed

mas@mp.pse.nagoya-u.ac.jp

## Abstract

A simple mechanical model was recently proposed to explain the universality of stretch at break ($\lambda_b$) as a function of cycle-rank density ($\xi$) in phantom-chain network simulations [J Non-Newtonian Fluid Mech., 349, 105620 (2026)]. Here, that model is reformulated as a series of the bottleneck strand and the surrounding network, yielding $\lambda_b - 1 = (\lambda_{bs} - 1)[1 + \nu_h/(1 + c\xi)]$. In this formula, $\lambda_{bs}$is the stretch at break of the bottleneck strand, $\nu_h$ is the number of stiff units in series along the rupture path, and $c$ is a geometric constant for the parallel redundancy of the medium. Since $c$ and $\nu_h$are difficult to separate over the examined range of $\xi$, $c$ is fixed, and $\lambda_{bs}$ and $\nu_h$are treated as fitting parameters. The formula is applied to phantom-chain simulations of networks with various conditions. In all cases, it reasonably captures the data, and the two parameters represent network characteristics.

## 1. Introduction

The fracture behavior of cross-linked polymer networks is sensitive to network topology [1–3] , yet a quantitative description linking molecular structure to macroscopic failure has remained elusive [4–10]. Recently, a series of simulation studies [11–17] has reported that the stretch at break $\lambda_b$ can be collapsed onto a universal master curve when plotted against the cycle rank density $\xi$, defined as

$$\xi = \frac{1}{n_n}(n_s - n_n) \qquad (1)$$

where $n_s$ and $n_n$ are the number of stress-supporting strands and nodes in the entire system, respectively [18–24]. This universality was first observed across phantom chain networks prepared from star prepolymers with monodisperse arm length and various arm numbers, $f$ and reaction conversions $p$ [11]. Similar $\xi$-dependence of $\lambda_b$ has been observed for other simulations, including phantom chain networks made from linear prepolymers and linkers [13], and bead-spring simulations with excluded volume interaction with random cross-linking [25].

A mechanical basis for this universality was provided in a companion paper [27]. The model replaces the network at the fracture with three springs in series: a single soft spring of stiffness $k_s$ (the bottleneck strand) flanked by two parallel bundles, each of stiffness $\eta k_s$. Identifying $\eta$ with $\xi$ through the unit-cell topology, with the three-element decomposition fixed, yields

$$\lambda_b - 1 = (\lambda_{bs} - 1)\,(3\xi + 6)/(3\xi + 2) \qquad (2)$$

where $\lambda_{bs}$ is a fitting parameter interpreted as the critical stretch of the bottleneck strand. Equation 2 was validated for Gaussian and FENE spring networks over a range of strand molecular weights $N_s$ [17], and a reasonable relation, $\lambda_{bs} \propto N_s^{1/2}$, was observed.

Equation 2 has only one fitting parameter $\lambda_{bs}$, which is essentially the shift-factor to normalize $(\lambda_b - 1)$ by $(\lambda_{bs} - 1)$. The $\lambda_b$-$\xi$ curves with various strand molecular weights in the previous study [27] behave in such a way, and the shape of the curves is essentially insensitive to $N_s$. However, as shown later, the phantom chain simulations with varying other parameters give $\lambda_b$-$\xi$ curves that cannot be fitted by eq 2, implying additional degrees of freedom. Independent support for this issue has recently been provided by Ishikura et al. [25], who confirmed a strong cycle-rank correlation with fracture stress and strain in their coarse-grained simulations, but argued that a second descriptor is needed to describe network fracture, identifying the effective chain ratio and strand-length uniformity as candidates.

In this study, the previously proposed model [27] is generalized. In the original model, each rupture path was assumed to consist of a fixed number of units. Here, this number is allowed to vary, while the construction is otherwise unchanged. The generalized model has a few parameters with direct structural meaning, into which the many parameters of the network are shown to reduce. The formula is applied to simulations whose $\xi$-dependence lies beyond eq 2, and the fitted parameters are shown to capture the structural characteristics of the networks. Details are shown below.

## 2. Model

Following the previous paper [27], this study focuses on the stretch at break $\lambda_b$. We consider the moment of rupture, at which the stress localizes to the single weakest strand (the bottleneck strand). The mechanical object of the model is not the system-spanning load path as a whole, but a local element: the bottleneck strand, together with the surrounding medium that connects it to the bulk on either side, as shown in Fig. 1. The bulk here refers to an effective elastic body that reproduces the mechanical response of the network far from the bottleneck. In this local view, $\lambda_b$ is set by the extension of the bottleneck and its immediate mechanical neighborhood, not by the length of any global path.

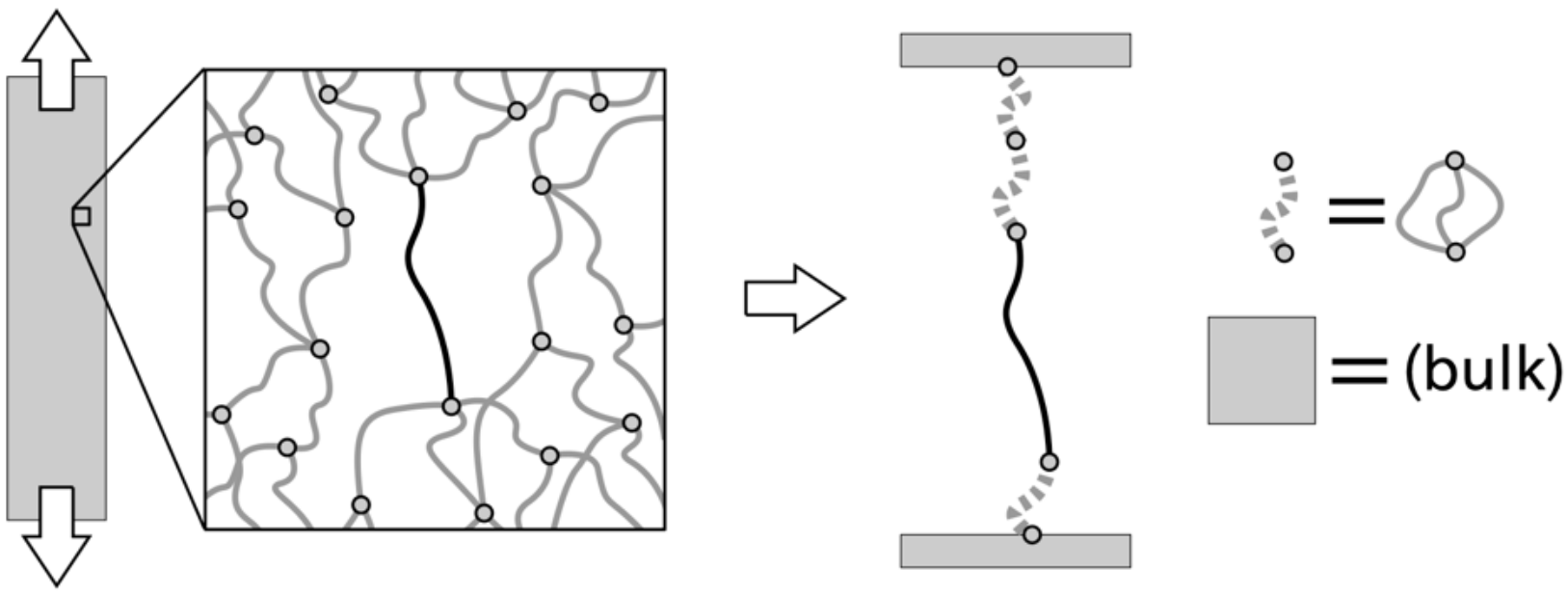

**Figure 1.** Schematic of the rupture path. A single soft bottleneck unit (the weakest strand, which ruptures) is in series with $\nu_h$ stiff units, each a bundle of $\eta$ parallel stress paths.

As sketched in Fig 1, the local element is represented as a series of mechanical units. The bottleneck strand is a single soft unit with stiffness $k_s$. On either side, the surrounding medium is represented by $\nu_h$ stiff units in series. Here $\nu_h$ is not the number of strands in the local region, but is the number of medium units over which the softness of the bottleneck response becomes indistinguishable from bulk. Each stiff unit has stiffness $\eta k_s$, where $\eta \geq 1$is the parallel redundancy of the medium: the load passing through the medium is shared among several parallel stress paths, so the medium is stiffer than a single strand.

For these units in series, the reciprocal path stiffness $k$ is

$$\frac{1}{k} = \frac{\nu_h}{\eta k_s} + \frac{1}{k_s} \qquad (3)$$

At rupture, the force reaches $F_b$, the force required to break the bottleneck strand, which is independent of the network. Thus the stretch at break of the entire system is

$$\lambda_b - 1 = \frac{F_b}{k} = \frac{F_b}{k_s}\left(\frac{\nu_h}{\eta} + 1\right) \qquad (4)$$

Here, lengths are measured in units of the rest length of the bottleneck strand, set to unity. Although the bottleneck spring itself has a microscopic natural length, its extension at rupture is expressed as a stretch relative to this reference length, consistent with the macroscopic $\lambda_b$. The prefactor $F_b/k_s$is thus the stretch of the bottleneck strand at rupture. Following the previous study, $F_b/k_s = (\lambda_{bs} - 1)$, and hence $\lambda_{bs}$is a strand-level quantity, expected to be independent of the network structure. Note that eq 4 is normalized by the bottleneck length, not by the total length $(\nu_h + 1)$ of the element. Namely, $\lambda_b$is the stretch of the whole element relative to the bottleneck strand, and it exceeds $\lambda_{bs}$ because the $\nu_h$medium units stretch together with the bottleneck.

The parallel redundancy $\eta$ is a property of the surrounding medium, and related to the cycle rank density $\xi$ of that medium. Note that the previous three-spring model [27] employed a fixed $\eta$ by counting strands and nodes within a small unit cell. Here, $\xi$ enters not as an internal count of the local element, but as a property of the effective medium in which the element is embedded. Although $\xi$ is a system-averaged quantity while the element is local, the two are linked through the effective medium, which is a representative part of the bulk network. Local variations of the cycle rank, such as those between $f =$ 3and $f = 8$ regions, are averaged out at this level, and the model holds in the mean-field sense.

A loop-free medium ($\xi = 0$) is a tree with no parallel redundancy, giving $\eta = 1$; each independent loop adds a parallel path. To lowest order, this gives a linear relation,

$$\eta = 1 + c\,\xi \qquad (5)$$

where $c$ is a geometric constant of order unity characterizing how the parallel redundancy of the medium builds up with cycle rank. Here, $\eta$ and $\nu_h$are effective descriptors of the medium, not integer counts, and are treated as continuous. Substituting eq 5 into eq 4 with $F_b/k_s = (\lambda_{bs} - 1)$ gives the central result,

$$\lambda_b - 1 = (\lambda_{bs} - 1)\left[1 + \frac{\nu_h}{1 + c\,\xi}\right] \qquad (6)$$

Equation 6 has a transparent structure. The constant term, $(\lambda_{bs}-1)$, is the stretch of the bottleneck itself, which is all that survives when the medium is stiff. The second term, $(\lambda_{bs}-1)\,\nu_h/(1+c\xi)$, is the additional stretch supplied by the surrounding medium; it decays as $\xi$ grows and the medium stiffens. The two limits are

$$\lambda_b - 1 \rightarrow \begin{cases} (\lambda_{bs}-1)(\nu_h+1) & \text{for } \xi \rightarrow 0 \\ (\lambda_{bs}-1) & \text{for } \xi \rightarrow \infty \end{cases} \qquad (7)$$

At $\xi \rightarrow 0$, the medium is a tree with no redundancy and contributes its full compliance, whereas at $\xi \rightarrow \infty$, the medium is rigid and only the bottleneck extends.

With $\nu_h = 2$ and $c = 3/2$, eq 6 reduces exactly to the three-spring formula of the previous study given by eq 2. The present formulation thus generalizes that work by treating $\nu_h$ and $c$ as structure-dependent quantities.

Although eq 6 nominally contains three parameters, $\lambda_{bs}$, $\nu_h$, and $c$, they are not all independently determinable from the available data. In the range accessible here, $\xi \lesssim 3$, a small-$\xi$ expansion of eq 6 gives

$$\lambda_b - 1 \simeq (\lambda_{bs}-1)(\nu_h+1) - (\lambda_{bs}-1)\,c\,\nu_h\,\xi + \cdots \qquad (8)$$

Eq 8 demonstrates that the initial slope is the product $(\lambda_{bs}-1)c\nu_h$; and thus, $c$ and $\nu_h$ are difficult to separate over the examined, limited range of $\xi$. Hereafter, $c$ is not fitted but fixed at a representative value, and $\lambda_{bs}$ and $\nu_h$ are treated as the two fitting parameters. In the previous three-spring model [27], $c = 3/2$ was implied; the present data are better described by $c = 4$, as shown below (Fig. 2). Therefore, $c$ is fixed at $c = 4$ throughout, except where noted. Note, however, that $c$ is not assumed universal. Its $\xi$-dependence, which requires data over a wider range of $\xi$, is left for future work.

## 3. Simulations

Since the simulation protocol follows previous studies [11,13–17,28,29], only a brief description is given here. Rouse–Ham-type star-branched prepolymers with $f$ arms of $N_a$ beads each were dispersed in a periodic simulation box with bead number density $\rho$. The beads are connected by the standard Gaussian linear springs. End-linking reactions were activated after equilibration, and primary loop formation was eliminated by disallowing intra-prepolymer reactions. The resulting network structures were stored at certain conversions $p$, and they were uniaxially stretched via sequential energy minimization and infinitesimal step-strain under volume-conservation conditions[29]. The bonds were severed when they exceeded the critical length $b_c$. The stretch at break $\lambda_b$ was taken from the peak of the nominal stress–strain curve [29], and $\xi$ was computed from Eq 1. Unless stated, $N_a = 5$, $\rho = 8$, ${b_c}^2 = 1.5$, $p$ was chosen at 0.6, 0.7, 0.8, 0.9, and 0.95, and $f$ was varied from 3 to 8. Eight independent realizations were performed for each network for statistics.

## 4. Results

Figure 2 (a) examines eq 6 against the dataset previously reported for the monodisperse star polymer networks with Gaussian springs [17]. Although $\nu_h = 2$ and $c = 3/2$ (hence eq 2) was used to describe the same dataset in the previous study [27], the fit shows better agreement with different values of $\nu_h$ and $c$, demonstrating the advantage of eq 6. In particular, the data in the range $\xi \lesssim 0.1$ is reproduced well if a larger $c$ value is chosen; see solid and broken curves, for which $c = 4$ and 3/2, respectively.

This effect of $c$ corresponds to the discussion in the previous paper about $\xi$-dependence of $\nu_h$, which is absorbed in $c$ in eq 6.

Figure 2 (b) shows the case of networks composed of $f = 3$ and $f = 8$ prepolymers, with the volume fraction of the $f = 3$ prepolymer, $\phi_3$, varied in the range $0 \leq \phi_3 \leq 1$. Although cycle-rank density is not uniform, as $f = 8$ polymers carry more effective paths than those with $f = 3$, a similar $\lambda_b$-$\xi$ curve is obtained, as reported earlier for FENE networks [16]. Further, eq 6 with the same parameter set as those in panel (a) describes the data, demonstrating that the effects of the mixing ratio and conversion are embedded into $\xi$, consistent with the model construction.

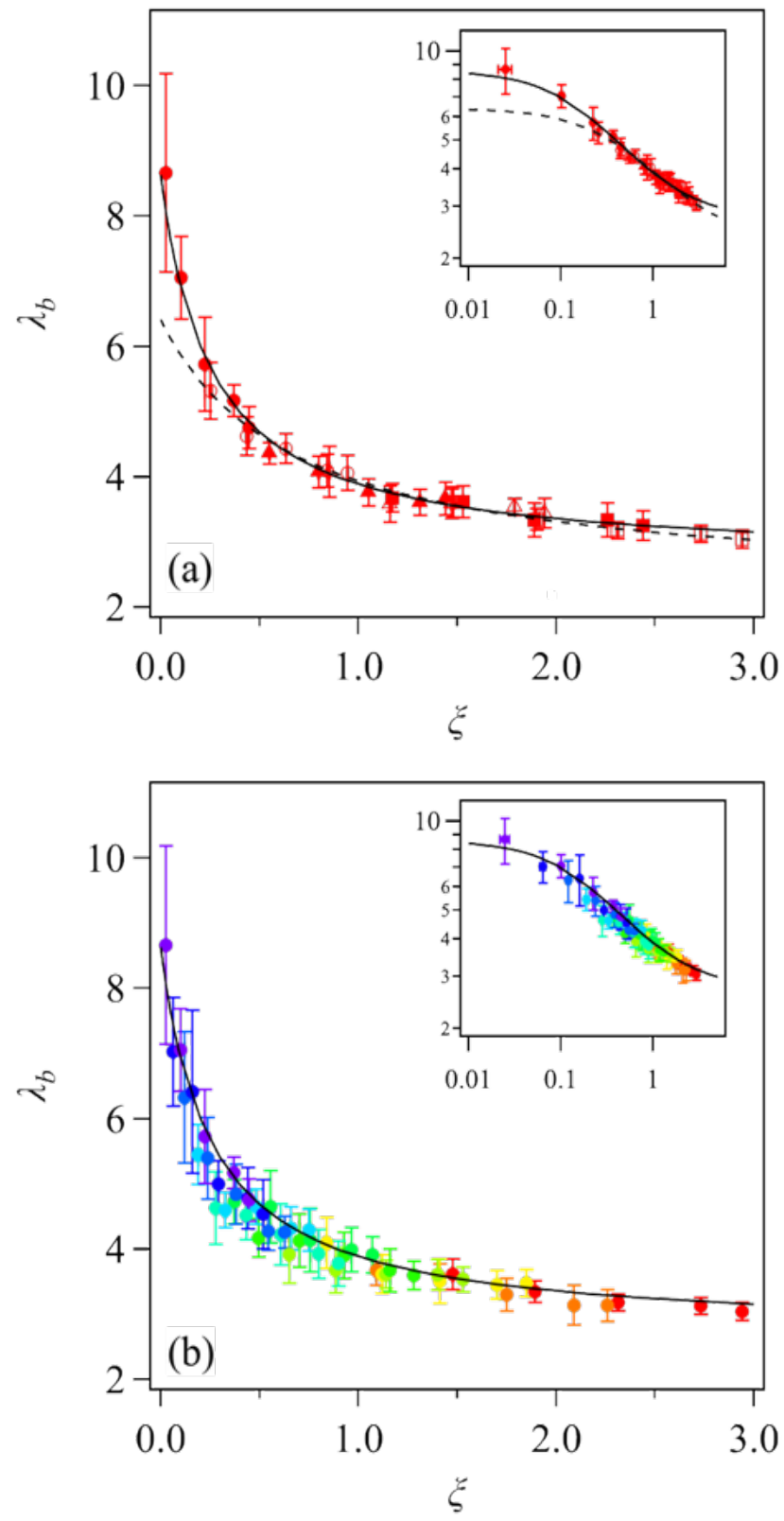

**Figure 2** $\lambda_b$-$\xi$ curves for the networks made from monodisperse star polymers with $f = 3\sim8$ [17] (a) and mixtures of $f = 3$ and 8 (b). Insets are logarithmic plots. In panel (a), symbols indicate $f = 3$ (filled circle), 4 (unfilled circle), 5 (filled triangle), 6 (unfilled triangle), 7 (filled square), and 8 (unfilled square). In panel (b), colors indicate the volume fraction of $f = 3$ polymers $\phi_3$ as 0 (red), 0.1 (orange), 0.2 (yellow), 0.3 (light green), 0.4 (green), 0.5 (sky blue), 0.6 (cyan), 0.7 (light blue), 0.8 (blue), 0.9 (dark blue), and 1.0 (violet). Error bars correspond to the standard deviations for eight different simulation runs for each condition. Solid and broken curves are fits to eq 6 with $\lambda_{bs} = 2.7 \pm 0.1$, $\nu_h = 3.6 \pm 0.4$, and $c = 4$, and $\lambda_{bs} = 2.3 \pm 0.1$, $\nu_h = 3.2 \pm 0.5$, and $c = 1.5$, respectively.

Figure 3 shows the results for various $b_c$ values. In this case, the initial network structures were common, and only the bond-breaking condition was modified. Consistent with this construction, the $\lambda_b$-$\xi$ curves in panel (a) are reasonably described by eq 6, irrespective of $b_c$. The $b_c$ −dependence of the model parameter $\lambda_{bs}$ in panel (b) shows a proportionality $\lambda_{bs} = (2.2 \pm 0.1)b_c$, as expected. Note that $\lambda_{bs}$ is not straightforwardly described as $2N_a b_c/b_0$, where $b_0$ is the average bond length before stretch, because bond extension is not uniform and bond scission occurs in elongated strands. Since the network structure is unaffected by $b_c$, $\nu_h$ should be independent of $b_c$, although $\nu_h$ slightly decreases as $b_c$ increases. This $b_c$-dependence of $\nu_h$ depends on the parameter $c$, which is fixed at 4 in this case.

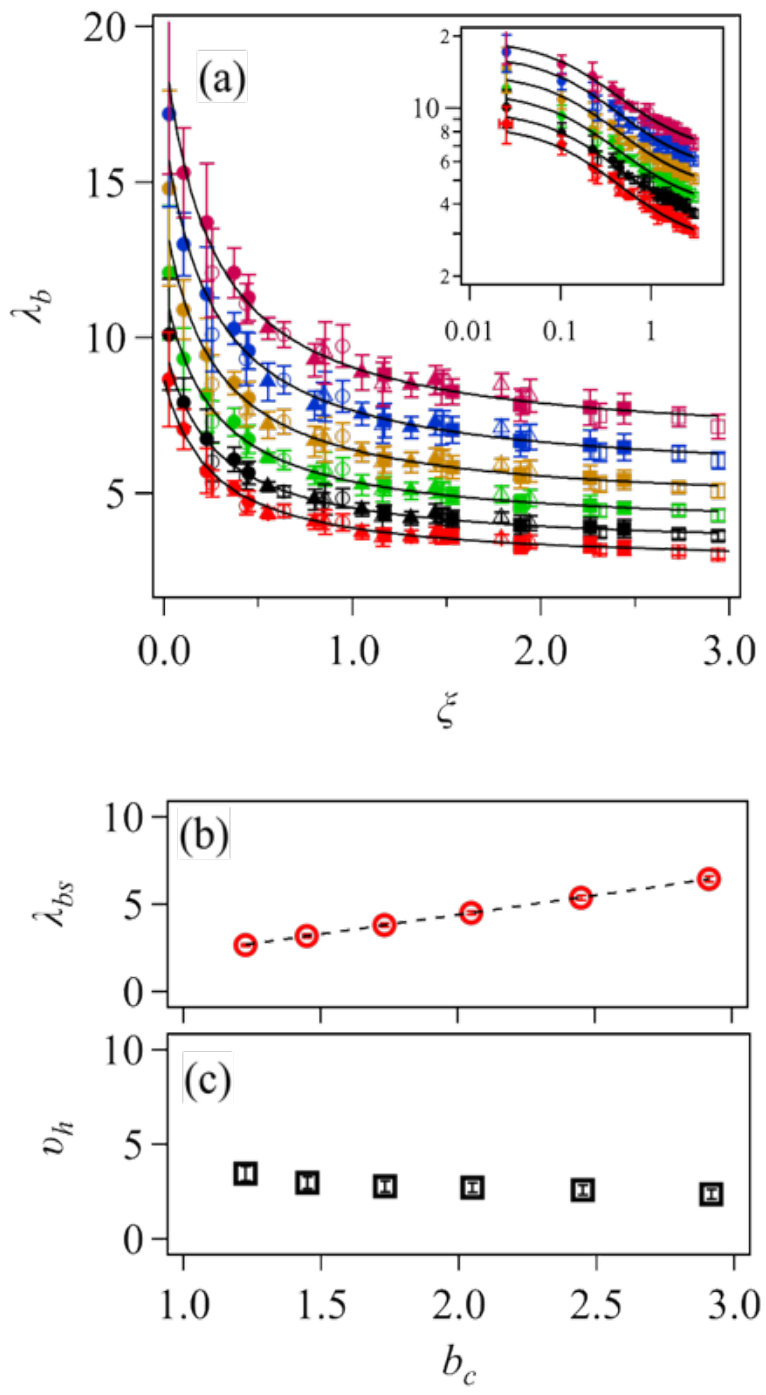

**Figure 3** $\lambda_b$-$\xi$ curves (a) for $b_c{}^2$=1.5 (red), 2.1 (black), 3 (green), 4.2 (brown), 6 (blue), and 8.5 (violet) with $f$ = 3 (filled circle), 4 (unfilled circle), 5 (filled triangle), 6 (unfilled triangle), 7 (filled square), and 8 (unfilled square). The inset shows a log-log plot, and error bars represent the standard deviations for eight different simulation runs for each condition. Solid curves indicate eq 6 with $c = 4$, and the parameters $\lambda_{bs}$ and $\nu_h$ are plotted against $b_c$ in panels (b) and (c). Error bars in panels (b) and (c) are fitting uncertainty. The broken line in panel (b) shows $\lambda_{bs} = 2.2b_c$.

Figure 4 shows the results for networks with different $\rho$ values. Although the networks are phantom, the rupture properties depend on $\rho$ due to differences in network connectivity induced by the end-linking reaction, as reported earlier for FENE spring networks [14]. Nevertheless, the $\lambda_b$-$\xi$ curves are reasonably described by eq 6 in panel (a) with the parameters shown in panels (b) and (c). In particular, $\nu_h$ decreases substantially as $\rho$ increases. This change of $\nu_h$ reflects the variation of $\lambda_b$-$\xi$ curve in its shape. Namely, the curve becomes flatter for large $\xi$ at larger $\rho$ values, and the $\xi$-dependence of $\lambda_b$ becomes weaker. According to the mechanical model in Fig. 1, the weak $\xi$-dependence is explained by a smaller value of $\nu_h$, which corresponds to a smaller series of stiff units and a stiffer path. Since $N_a$ and $b_c$ are common, $\lambda_{bs}$ was expected to be independent of $\rho$, but it slightly increases with increasing $\rho$. This increase in $\lambda_{bs}$ may reflect a change in network connectivity, but it depends on the parameter $c$, which is fixed at 4, as in Fig. 3.

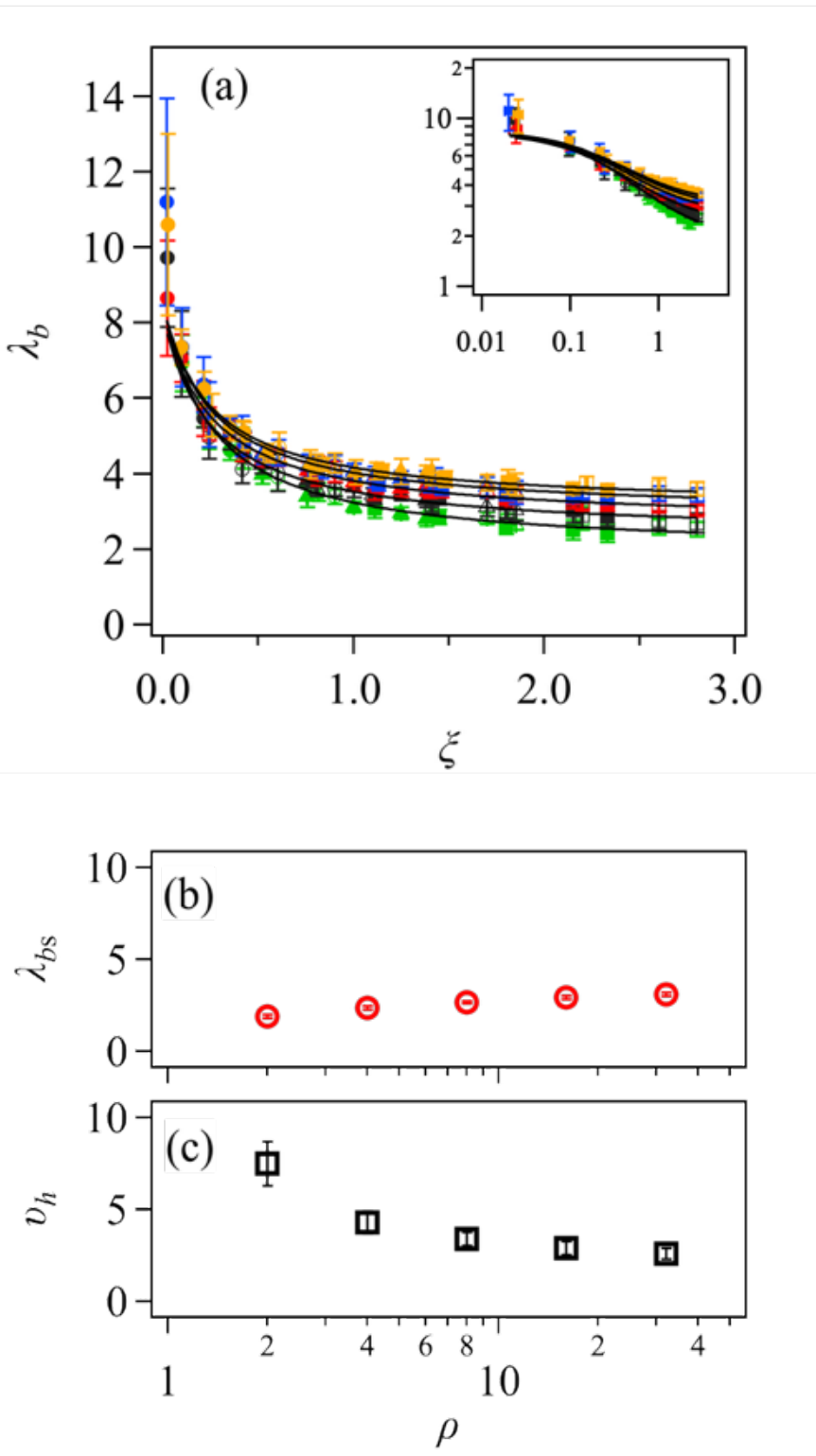

**Figure 4** $\lambda_b$-$\xi$ curves (a) for $\rho = 2$ (green), 4 (black), 8 (red), 16 (blue), and 32 (yellow) with $f = 3$ (filled circle), 4 (unfilled circle), 5 (filled triangle), 6 (unfilled triangle), 7 (filled square), and 8 (unfilled square). The inset shows a log-log plot, and error bars represent the standard deviations for eight different simulation runs for each condition. Solid curves indicate eq 6 with $c = 4$, and the parameters $\lambda_{bs}$ and $\nu_h$ are plotted against $\rho$ in panels (b) and (c). Error bars in panels (b) and (c) are fitting uncertainty.

Figure 5 shows the case with various $N_a$, for which the dataset was reported earlier[17]. $\lambda_b$-$\xi$curves in panel (a) are reasonably described by eq 6 with the parameters shown in panel (b). Concerning the fitting parameters, $\lambda_{bs}$ depends on the strand bead number $N_s = 2N_a + 1$ as $\lambda_{bs} \propto {N_s}^{0.65}$. The exponent is similar to that reported in earlier studies on the molecular-weight dependence of $\lambda_b$ [17]. Note, however, that $\lambda_{bs}$ is different from $\lambda_b$; the latter includes the effects of $\xi$ and $\nu_h$. The power-law exponent differs if $N_s$-dependence of $\lambda_b$ is evaluated at a certain set of $f$ and $p$. Note also that the data show slight deviations from a single power law at the smallest and largest $N_s$, so the exponent should be regarded as an effective, approximate value. Concerning $\nu_h$, differently from the case with varying $b_c$, $\nu_h$ strongly decreases with increasing $N_s$, similar to that against $\rho$. This result is due to the fact that the prepolymer concentration with respect to the overlapping concentration, $c/c^*$, increases as $N_a$ increases since the bead number density $\rho$ is fixed at 8.

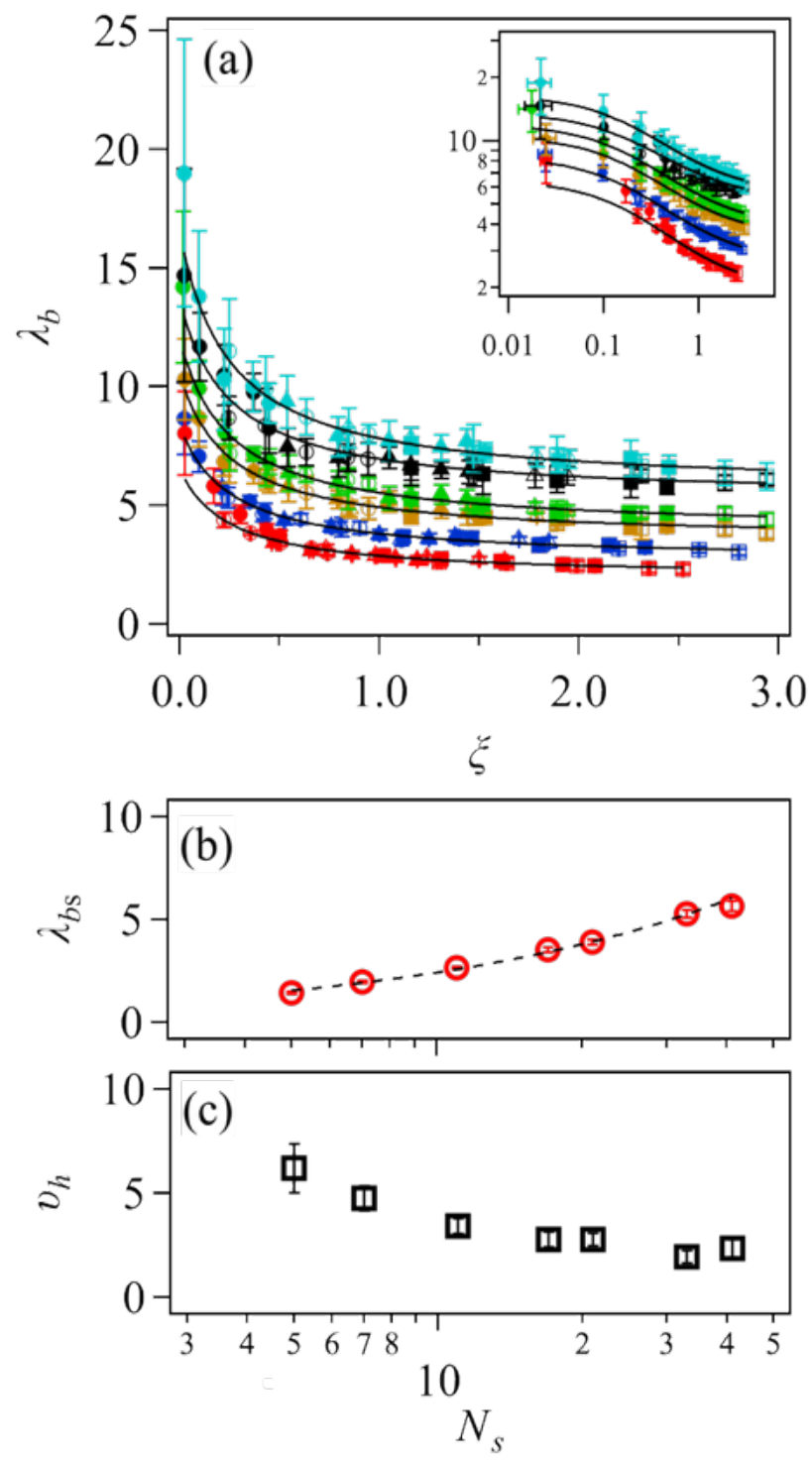

**Figure 5** $\lambda_b$-$\xi$ curves (a) for $N_a$=3 (red), 5 (blue), 8 (yellow), 10 (green), 16 (black), and 32 (sky blue) with $f$ = 3 (filled circle), 4 (unfilled circle), 5 (filled triangle), 6 (unfilled triangle), 7 (filled square), and 8 (unfilled square). The inset shows a log-log plot, and error bars represent the standard deviations for eight different simulation runs for each condition. Solid curves indicate eq 6 with $c = 4$, and the parameters $\lambda_{bs}$ and $\nu_h$ are plotted against the strand bead number $N_s = 2N_a + 1$ in panels (b) and (c). Error bars in panels (b) and (c) are fitting uncertainty. The broken curve in panel (b) shows $\lambda_{bs} = 0.5{N_s}^{0.65}$.

## 5. Discussion

The path stiffness $k$ discussed above should be distinguished from the macroscopic modulus $G$ and $E$. At small strain, the network responds collectively, and $G$ scales with the cycle rank $\xi$ as in phantom network theory [30–32]. As the strain approaches $\lambda_b$, the stress localizes onto the weakest path containing the bottleneck strand, and the rupture is governed by the stiffness $k$ of that single path. Such progressive redistribution of stress toward the weakest strands during deformation has been documented [33–39]. The present model describes this rupture limit; accordingly, $k$ remains finite as $\xi$ approaches zero (in contrast to $G \to 0$ as $\xi \to 0$), because the two characterize different stages of deformation. Converting $k$ (force per length) to a stress, such as $G$ requires the areal density of bottleneck paths, a spatial quantity that depends on the network structure. Obtaining the fracture stress or fracture energy would require this density, and a full stress–strain description would further require how the density of load-bearing paths evolves with strain. The present model deliberately sets these aside and is formulated purely in terms of network topology, the number of stiff units in series ($\nu_h$), and the cycle rank ($\xi$). Because $\lambda_b$ is a dimensionless stretch, it is determined solely by this topology, whereas a quantitative prediction of the stress requires the spatial path density. The present model isolates the topological content of $\lambda_b$ and thereby reveals its universality in $\xi$. The distinction between $k$ and $G$ also clarifies an earlier observation: the nonaffine deformation theory of Panyukov and Rubinstein [40] was found not to account for the strand-length dependence of $\lambda_b$. The nonaffine theory describes the average, collective deformation of the network, corresponding to the small-strain regime dominated by $G$, whereas $\lambda_b$ is governed by $k$ for the single weakest path at rupture. Accordingly, the present description applies to the stress–strain response near rupture under uniaxial extension, where the load localizes onto the weakest path. It is not intended for the small-strain moduli $G$ and $E$, to which deformation modes other than uniaxial extension also contribute.

The local element of Fig. 1 should not be read as a system-spanning path, but it is the bottleneck strand and its immediate mechanical surroundings. The small fitted values of $\nu_h$(typically a few) are consistent with this view: the softness of the bottleneck is screened by the surrounding medium over a short effective range, beyond which the response is bulk-like. Correspondingly, the cycle rank $\xi$ enters eq 6 as a property of that embedding medium, not as a count internal to the element. This construction resolves the apparent tension between a global, sample-spanning $\xi$ and a local mechanical object: they are connected because the medium surrounding the bottleneck is the part of the bulk network characterized by $\xi$.

## 6. Concluding Remarks

A mechanical model for the cycle rank density $\xi$ dependence of the stretch at rupture $\lambda_b$, found in phantom chain networks, was presented as a generalization of the previous three-spring description. Modeling the rupture path as stiff units in series with a single soft bottleneck unit, and fixing the parallelism through the cycle rank, gives the formula eq 6, $\lambda_b - 1 = (\lambda_{bs} - 1)[1 + \nu_h/(1 + c\xi)]$. It contains three parameters: $\lambda_{bs}$, the stretch at break of the bottleneck strand; $\nu_h$, the number of stiff units in series per rupture path; and $c$, the geometric constant for the parallel redundancy of the medium. Since $c$ and $\nu_h$are difficult to separate over the examined range of $\xi$, $c$ is fixed, and the many structural parameters of the network reduce to the two fitting parameters $\lambda_{bs}$ and $\nu_h$.

Although the fundamental idea of this study, to describe the $\lambda_b$-$\xi$ curve from the stiffness of the bottleneck path, works reasonably well, several questions are left for future work. For instance, the parameters $\lambda_{bs}$ and $\nu_h$ do not fully serve their intended roles as strand-level and structural descriptors. A complete, microscopically grounded separation is required. The parameter $c$, which was set as a constant in this work, should be discussed by fitting datasets spanning wider ranges of $\xi$. Most importantly, the crossover from the small-strain behavior dominated by the modulus $G$, to the rupture governed by the path stiffness $k$, is non-trivial, and it probably links this study to the other earlier studies on the shortest path and the elastic backbone ideas. Finally, experimental verification of eq 6 is necessary. Studies toward such directions are ongoing, and the results will be reported elsewhere.

### Acknowledgements

This study was partly supported by the Hibi Foundation and JSPS KAKENHI (26H02291).